\documentclass[final]{aa}

\usepackage{graphicx}
\usepackage{natbib}

\begin{document}

\title{Fine-structure diagnostics of neutral carbon
  toward~\object{HE~0515-4414}
  \thanks{Based on observations made with ESO Telescopes at the La Silla
    or Paranal Observatories under programme ID~066.A-0212}}
\author{Ralf Quast \and
  Robert Baade \and
  Dieter Reimers}
\authorrunning{R.~Quast et al.}
\institute{Hamburger Sternwarte, Universit\"at Hamburg,
  Gojenbergsweg 112, D-21029 Hamburg, Germany \\
  \email{[rquast,rbaade,dreimers]@hs.uni-hamburg.de}}
\offprints{R.~Quast}
\date{Received \today / Accepted ?}

\abstract{%
New high-resolution high signal-to-noise spectra of the $z=1.15$ damped
Lyman~$\alpha$ (DLA) system toward the quasi-stellar object HE~0515-4414
reveal absorption lines of the multiplets~2 and 3 in \ion{C}{i}. The
resonance lines are seen in two components with total column densities of
$\log N=13.79\pm0.01$ and $\log N=13.36\pm0.01$, respectively. The
comparision of theoretical calculations of the relative fine-structure
population with the ratios of the observed column densities suggests
that the \ion{C}{i} absorbing medium is either very dense or exposed to
very intense UV radiation. The upper limit on the local UV energy density
is 100 times the galactic UV energy density, while the upper limit on the
\ion{H}{i} number density is 110~cm$^{-3}$. The excitation temperatures of
the ground state fine-structure levels of $T=15.7$ and $T=11.1$~K,
respectively, are consistent with the temperature-redshift relation
predicted by the standard Friedmann cosmology. The cosmic microwave
background radiation (CMBR) is only a minor source of the observed
fine-structure excitation.

\keywords{cosmic microwave background --
  intergalactic medium --
  quasars: absorption lines --
  quasars: individual: HE~0515-4414}}

\maketitle

\section{Introduction} 

The standard Friedmann cosmology predicts that the temperature of the
cosmic microwave background radiation (CMBR) increases linearly with the
redshift $z$ as
\begin{equation}
  T_{\mathrm{CMBR}}(z) =
    T_{\mathrm{CMBR}}(0)\,(1+z).
  \label{eq:TCMBR}
\end{equation}
The present-day CMBR temperature has been ascertained by the Cosmic
Background Explorer FIRAS instrument to be $T_{\mathrm{CMBR}}(0)=2.725$~K
\citep{JMatherFSMW1999}. In recent years several attempts have been made
to infer the CMBR temperature at higher redshifts from the relative
population of \ion{C}{i} (and \ion{C}{ii}) fine-structure levels observed
in damped Lyman~$\alpha$ (DLA) systems \citep{JGeBB1997,JGeBK2001,
KRothJBauer1999,RSrianandPL2000}. The principal problem in this inference
is the presence of additional sources of excitation. In particular,
collisional excitation and fluorescence induced by the local UV radiation
field are competing processes. A recent study of \citet{ASilvaSViegas2002}
shows the possibility to assess the physical conditions in QSO absorbers
from the observation of fine-structure absorption lines.

This study provides a fine-structure diagnostics of \ion{C}{i} toward the
quasi-stellar object HE~0515-4414 ($z=1.73$, $B=15.0$) discovered by the
Ham\-burg/ESO Survey \citep{DReimersHRW1998}. The observed fine-structure
absorption lines (and a large number of additional metal lines) are
associated with a DLA system at $z=1.15$ \citep{AVargaRTBB2000}. We infer
the physical conditions in the DLA system and demonstrate that the CMBR is
only a minor source of the observed fine-structure excitation.

\section{Observations} 

HE~0515-4414 was observed during ten nights between October~7, 2000 and
January~3, 2001, using the UV-Visual Echelle Spectrograph (UVES)
installed at the second VLT Unit Telescope (Kueyen). Thirteen exposures
were made in the dichroic mode using standard settings for the central
wavelenghts of 3460/4370~\AA\ in the blue, and 5800/8600~\AA\ in the
red. The CCDs were read out in fast mode without binning. Individual
exposure times were 3600 and 4500 seconds, under photometric to clear
sky and seeing conditions ranging from 0.47 to 0.70 arcseconds. The slit
width was 0.8 arcseconds resulting in a spectral resolution of about
55\,000 in the blue and slightly less in the red.

The raw frames were reduced at Quality Control Gar\-ching using the UVES
pipeline Data Reduction Software \citep{PBallesterMBC2000}. Finally, the
individual vacuum-barycentric corrected spectra were co-added resulting in
an effective signal-to-noise ratio typically better than 100 (up to 130
for the parts of the spectrum considered in this study).

\begin{figure*}
  \sidecaption
  \resizebox{12cm}{!}{\includegraphics*[94pt,13pt][435pt,425pt]{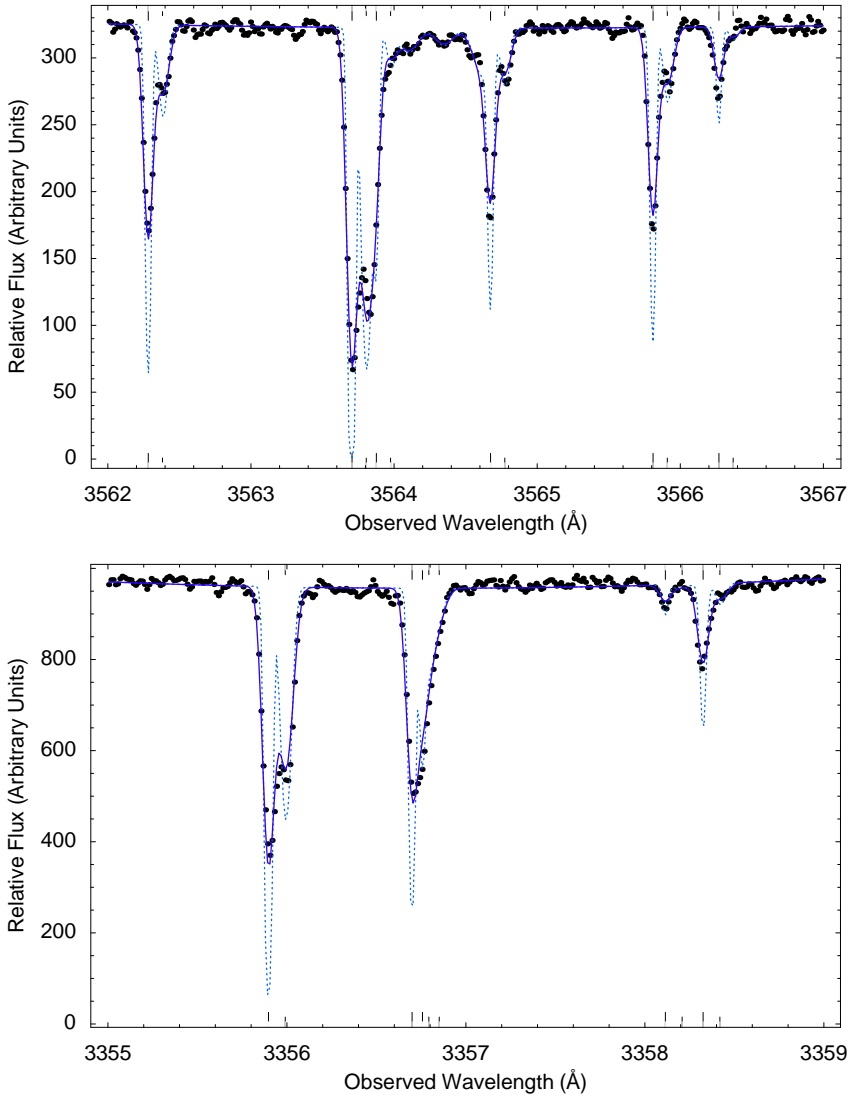}}
  \caption{~Parts of the spectrum showing the lines of the multiplets
    2 (top panel) and 3 (bottom panel) in \ion{C}{i}. The fat dots
    represent the observed flux. The solid and dashed curves represent
    our optimized model and its deconvolution, respectively. The long
    (short) vertical tick marks indicate the major (minor) absorption
    component in each composite profile. The effective signal-to-noise
    ratio between the $^3$P$_1$--$^3$P$_1$ and $^3$P$_1$--$^3$P$_0$ lines
    (top panel) is about 130. The standard error provided with the reduced
    flux data is about twice the noise}

  \label{fg:spectrum}
\end{figure*}

\section{Absorption line analysis} 

The observed spectral flux $F$ is the convolution of the instrumental
profile $P$ with the product of the background continuum $C$ and the
absorption term:
\begin{equation}
  F(\lambda) =
    \int P(\lambda')C(\lambda-\lambda')
      \exp{[-\tau(\lambda-\lambda')]}\,\mathrm{d}\lambda'.
  \label{eq:F}
\end{equation}
The instrumental profile is approximated by a normalized Gaussian defined
by the spectral resolution of the instrument. The background continuum
is locally approximated by a linear combination of Legendre polynomials
of up to second order. Assuming pure Doppler broadening, the optical depth
is modeled by a superposition of Gaussian functions:
\begin{equation}
  \tau(\lambda) =
    \sum_i g_i(\lambda),
  \label{eq:tau}
\end{equation}
where
\begin{equation}
  g_i(\lambda) =
    \frac{e^2}{4\varepsilon_0 mc}
    \frac{N_i f_i \lambda_i}{\sqrt{\pi}b_i}\,
      \exp{\!\left[-\left(c\,
        \frac{\lambda/(1+z_i)-\lambda_i}{\lambda_i b_i}\right)^2
        \right]}
  \label{eq:g}
\end{equation}
with $\lambda_i$, $f_i$, $z_i$, $b_i$ and $N_i$ denoting the rest
wavelength, the oscillator strength, the redshift, the line broadening
velocity and the column density of a single line, respectively. Each
\ion{C}{i} absorption component is modeled by a superposition of Doppler
profiles with identical redshifts and widths. Additionally, the particular
lines originating on the same fine-structure level are restricted to have
identical column densities. This simultaneous fine-structure treatment
ensures the physical consistency of our analysis. The weak absorption
features between the $^3$P$_2$--$^3$P$_2$ and $^3$P$_1$--$^3$P$_1$ lines
seen in Fig.~\ref{fg:spectrum} are modeled by a superposition of three
artificial Doppler profiles.

Our model is defined by 24 strongly correlated parameters: six
parameters to model the background continuum, 17 parameters to model the
optical depth, and the spectral resolution of the instrument. In order
to find the optimal set of parameter values, we minimize the $\chi^2$
statistic following an adaptive evolution strategy recently proposed by
\citet{NHansenAOstermeier2001}.

The atomic line data used in this analysis are collected in
Table~\ref{tb:adata}.

\begin{table}
  \caption{Atomic data for the radiative transitions of the multiplets
    2 and 3 in \ion{C}{i}. The vacuum wavelengths $\lambda$ are taken from
    \citet{DMorton1991}, the oscillator strengths $f$ are adopted from the
    compilation of \citet{WWieseFD1996}}

  \begin{tabular}{lllll}
    \hline
    Mult.\rule[-6pt]{0pt}{18pt}
      & \multicolumn{2}{l}{Transition}
                               & $\lambda$ (\AA) & $f$ (10$^{-2}$) \\
    \hline
    2\rule{0pt}{12pt} & 2p$^2$--
        & $^3$P$_1$--$^3$P$_2$ & 1656.2672       & 5.89 \\
      & \phantom{2}2p\,3s
        & $^3$P$_0$--$^3$P$_1$ & 1656.9283       & 13.9 \\
      & & $^3$P$_2$--$^3$P$_2$ & 1657.0082       & 10.4 \\
      & & $^3$P$_1$--$^3$P$_1$ & 1657.3792       & 3.56 \\
      & & $^3$P$_1$--$^3$P$_0$ & 1657.9068       & 4.73 \\
      & & $^3$P$_2$--$^3$P$_1$ & 1658.1212       & 3.56 \\
    3 & 2s$^2$2p$^2$--
        & $^3$P$_0$--$^3$D$_1$ & 1560.3092       & 7.19 \\
      & \phantom{2}2s\,2p$^3$
        & $^3$P$_1$--$^3$D$_2$ & 1560.6822       & 5.39 \\
      & & $^3$P$_1$--$^3$D$_1$ & 1560.7090       & 1.80 \\
      & & $^3$P$_2$--$^3$D$_2$ & 1561.3402       & 1.08 \\
      & & $^3$P$_2$--$^3$D$_1$ & 1561.3667       & 0.07 \\
      & & $^3$P$_2$--$^3$D$_3$ & 1561.4384       & 6.03 \\
    \hline
  \end{tabular}

  \label{tb:adata}
\end{table}

\section{Results and discussion} 

\subsection{Model parameters}

The optimized values and formal confidence limits of the model parameters
for both \ion{C}{i} absorption components are collected in
Table~\ref{tb:pdata}. The total column density of \ion{C}{i} is $\log
N=13.79\pm0.01$ for the major and $\log N =13.36\pm0.01$ for the minor
component. These values are consistent with the total column density $\log
N=13.90\pm0.04$ reported by \citet{AVargaRTBB2000} for the non-separated
\ion{C}{i} absorption components.

\begin{table}
  \caption{Optimized values and formal 68.3 percent confidence limits of
    the model parameters for both \ion{C}{i} absorption components. The
    redshift is accurate to the last digit}

  \begin{tabular}{llll}
    \hline
    Level\rule[-6pt]{0pt}{18pt}
      & $z$       & $b$ (km\,s$^{-1}$)   & $\log N$ (cm$^{-2}$) \\
    \hline
    $^3$P$_0$\rule{0pt}{12pt}
      & 1.150789  & $2.00\pm0.03$        & $13.53\pm0.01$ \\
    $^3$P$_1$
      &           &                      & $13.35\pm0.01$ \\
    $^3$P$_2$
      &           &                      & $12.76\pm0.02$ \\
    $^3$P$_0$
      & 1.150853  & $3.53\pm0.12$        & $13.21\pm0.01$ \\
    $^3$P$_1$
      &           &                      & $12.76\pm0.02$ \\
    $^3$P$_2$
      &           &                      & $12.10\pm0.07$ \\
    \hline
  \end{tabular}

  \label{tb:pdata}
\end{table}

Figure~\ref{fg:spectrum} reveals a saturated narrow structure in the
$^3$P$_0$ line profiles of the major absorption component. This problem
can be tackled indirectly by correcting the apparent optical depth
\citep{BSavageKSembach1991,EJenkins1996}, but the correction procedure is
established for particular synthetic data only and may not be applicable
in general. Instead, we tackle the problem of saturated narrow structure
directly by testing our analytical procedure with synthetic data similar
to the observed spectrum. The tests confirm that our analytical procedure
correctly recovers the narrow structure in the $^3$P$_0$ line profiles if
the saturation is moderate. Presuming stronger saturation gives rise to
observed fine-structure populations being far away from statistical
equilibrium.

The optimized value of the spectral resolution of the instrument
$R=54\,600\pm800$ matches the spectral resolution of the individual
exposures.

\subsection{Physical conditions}

The ground state of the carbon atom consists of the 2s$^2$2p$^2$
$^3$P$_{0,1,2}$ fine-structure triplet levels. In DLA systems, the
excited levels of the ground state triplet are populated principally by
three competing processes: direct photoexcitation by the CMBR,
fluorescence induced by the local UV radiation field, and collisional
excitation by hydrogen atoms \citep{JBahcallRWolf1968}. Given the
physical conditions, the relative population of fine-structure levels is
determined by solving the system of statistical equilibrium equations. In
order to calculate the solution, we use the PopRatio program package
recently developed by \citet{ASilvaSViegas2001}. The package provides a
Fortran~90 source code and an up to date compilation of atomic data for
\ion{C}{i} (and \ion{C}{ii}, \ion{O}{i}, \ion{Si}{ii}).

The equilibrium equations are established presuming the CMBR temperature
follows Eq.~(\ref{eq:TCMBR}) and the local UV input equals the scaled
generic galactic radiation field. The direct photoexcitation rates by the
CMBR at $z=1.15$ are $R_{01}=4.3\,10^{-9}$~s$^{-1}$ and
$R_{02}=2.5\,10^{-18}$~s$^{-1}$, while fluorescence (the PopRatio package
considers 108 UV transitions) induced by the generic galactic UV radiation
field \citep{PGondhalekarPW1980} yields UV pumping rates of
$R_{01}=3.5\,10^{-10}$~s$^{-1}$ and $R_{02}=2.8\,10^{-10}$~s$^{-1}$. The
\ion{H}{i} collision rates at the kinetic temperature
$T_{\mathrm{kin}}=100$~K are $q_{01}=3.8\,10^{-10}$~cm$^3$\,s$^{-1}$ and
$q_{02}=2.5\,10^{-10}$~cm$^3$\,s$^{-1}$. We point out that the UV pumping
rates will exceed the direct photoexcitation rates if the local UV input
exceeds the generic galactic radiation field by more than a factor of
twelve.

If the absorbing medium is homogenous, the relative population of
excited and ground fine-structure levels matches the corresponding
column density ratios, $X_{1,2}=N_{1,2}/N_0$. Consequently, if
the column densities $N_{0,1,2}$ are regarded as independent random
observables with density functions $p_{0,1,2}(N_{0,1,2})$, $X_{1,2}$ are
random observables with cumulative distribution functions
\begin{equation}
  F(X_{1,2}) =
    \int_0^\infty\int_0^{N_0X_{1,2}}
      p_0(N_0)p_{1,2}(N_{1,2})
        \,\mathrm{d}N_{1,2} \,\mathrm{d}N_0.
  \label{eq:CDF}
\end{equation}
Deriving the cumulative distribution functions with respect to $X_{1,2}$
yields the density functions
\begin{equation}
  p(X_{1,2}) =
    \int_0^\infty
      p_0(N_0)p_{1,2}(N_0X_{1,2})N_0 \,\mathrm{d}N_0.
  \label{eq:p}
\end{equation}

\begin{figure}
  \centering
  \resizebox{\hsize}{!}{\includegraphics*[94pt,23pt][344pt,234pt]{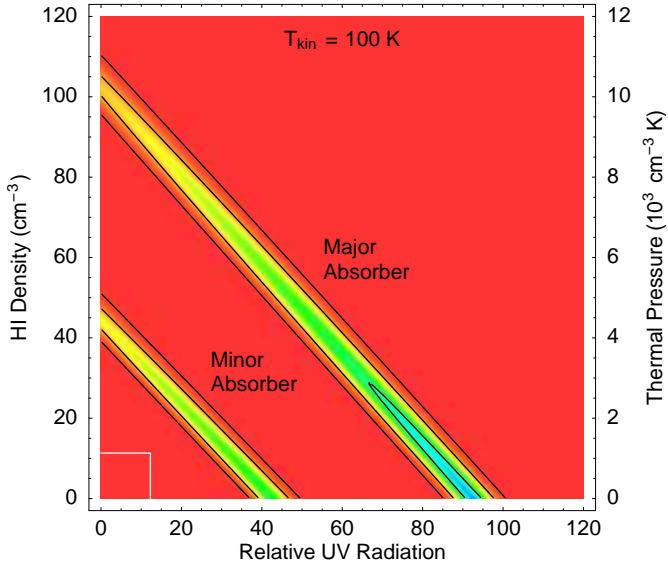}}
  \caption{Probability density of the physical conditions in the DLA
    system. The contour lines are drawn at 0.61, 0.14, and 0.01 of the
    maximum and would correspond to the boundaries of the 68.3, 95.4,
    and 99.7 percent confidence regions if the distributions were normal.
    The rectangle in the lower left corner marks the physical conditions
    when the CMBR would be the principal source of the fine-structure
    excitation}

  \label{fg:phys100}
\end{figure}

\begin{figure}
  \centering
  \resizebox{\hsize}{!}{\includegraphics*[94pt,23pt][344pt,234pt]{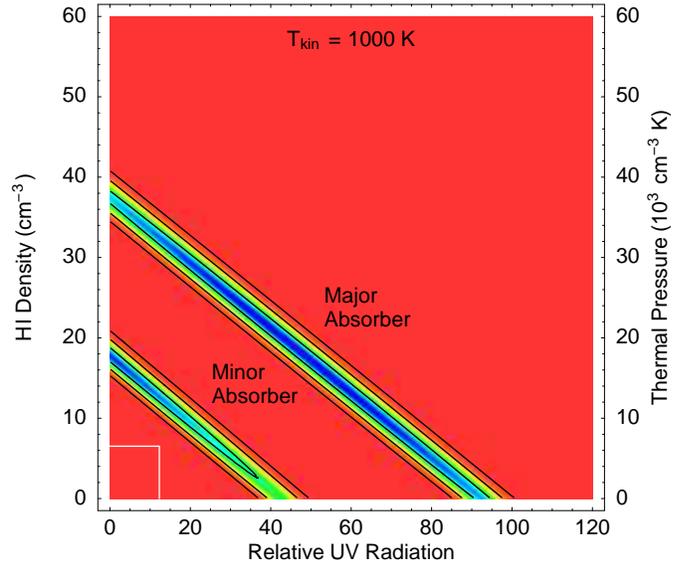}}
  \caption{The same as Fig.~\ref{fg:phys100}, but considering a ten times
    higher kinetic temperature. Note the different scale of the vertical
    axis}

  \label{fg:phys1000}
\end{figure}

In order to infer the physical conditions in each observed \ion{C}{i}
absorber, we define $p_{0,1,2}(N_{0,1,2})$ by the requirement that
$\log N_{0,1,2}$ are normally distributed with mean values and standard
deviations matching the optimized values collected in
Table~\ref{tb:pdata}. Then, we calculate the equilibrium population
$X_{1,2}$ for a grid of physical conditions and evaluate the joint
density function
\begin{equation}
  p(X_1,X_2) =
    p(X_1)p(X_2).
  \label{eq:p12}
\end{equation}
Contour diagrams of the joint density functions obtained in this way are
shown in Fig.~\ref{fg:phys100}, considering a kinetic temperature of
$T_{\mathrm{kin}}=100$~K. Figure~\ref{fg:phys100} clearly demonstrates
that the CMBR is only a minor source of the observed fine-structure
excitation. Considering the 0.01 contour line (i.e. the boundary of the
99.7 percent confidence region if the distribution were normal), the upper
limit on the UV energy density in the major (minor) \ion{C}{i} absorber is
100 (50) times the galactic UV energy density. The upper limit on the
\ion{H}{i} number density is 110~cm$^{-3}$ (50~cm$^{-3}$). The upper limit
on the thermal pressure of 11\,000~cm$^{-3}$\,K (5000~cm$^{-3}$\,K)
exceeds the median pressure in galactic \ion{C}{i} absorbers
\citep{EJenkinsTTripp2001} by a factor of four (two). A similar analysis
considering a kinetic temperature of $T_{\mathrm{kin}}=1000$~K (see
Fig.~\ref{fg:phys1000}) yields upper limits of 41~cm$^{-3}$ (21~cm$^{-3}$)
and 41\,000~cm$^{-3}$\,K (21\,000~cm$^{-3}$\,K). If the CMBR is completely
excluded from the analysis, the limits increase by less than 20~percent.

\begin{figure}
  \centering
  \resizebox{\hsize}{!}{\includegraphics*[94pt,23pt][344pt,234pt]{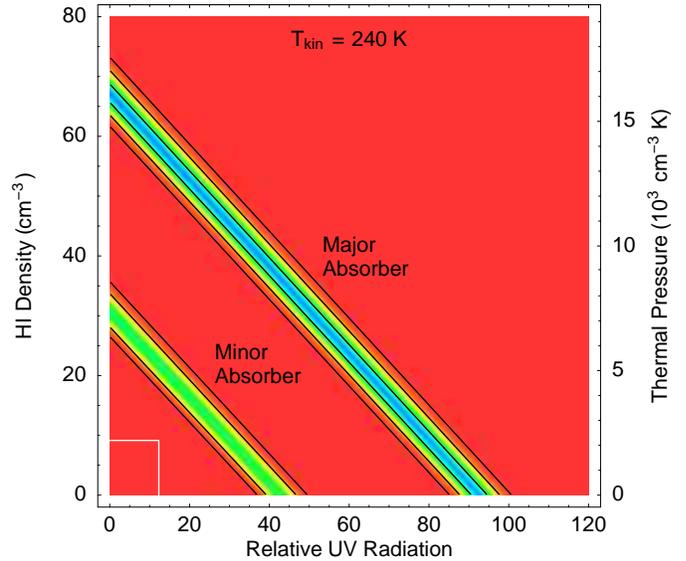}}
  \caption{The same as Figs.~\ref{fg:phys100} and~\ref{fg:phys1000},
    but considering the most probable kinetic temperature. Note the
    different scale of the vertical axis}

  \label{fg:phys240}
\end{figure}

If we integrate the joint density function with respect to the \ion{H}{i}
number density and the relative UV energy density, we obtain the marginal
probability density of the kinetic temperature. The marginal density
function cannot rule out any kinetic temperature in the range
$\mathrm{40~K}\le T_{\mathrm{kin}}\le\mathrm{1000~K}$, but indicates that
in both \ion{C}{i} absorbers a kinetic temperature of about
$T_{\mathrm{kin}}=240$~K is the most probable. The corresponding contour
plot is shown in Fig.~\ref{fg:phys240}.

\subsection{Excitation temperature}

If the absorbing medium is homogenous, the excitation temperature is
defined (via the Boltzmann equation) by the ratio of column densities
\begin{equation}
  \frac{N_j}{N_i} =
    \frac{g_j}{g_i}\, \exp\left[-\frac{E_{ij}}{kT_{ij}}\right].
  \label{eq:Tij}
\end{equation}
The energies of the first and second excited ground state fine-structure
levels in \ion{C}{i} relative to the ground level are
$E_{01}=16.4$~cm$^{-1}$ and $E_{02}=43.4$~cm$^{-1}$. Now considering the
optimized column density values listed in Table~\ref{tb:pdata}, we
obtain three different excitation temperatures for each \ion{C}{i}
absorber (see Table~\ref{tb:tdata}). These excitation temperatures are
considerably higher than the CMBR temperature predicted by
Eq.~(\ref{eq:TCMBR}), $T_{\mathrm{CMBR}}(1.15)=5.9$~K. The magnitude and
diversity of the excitation temperatures reveals that the CMBR is not the
principal source of the fine-structure excitation. Moreover, the
excitation temperatures in the minor absorber are systematically lower,
indicating completely different physical conditions.

\begin{figure}
  \centering
  \resizebox{\hsize}{!}{\includegraphics*[94pt,4pt][344pt,158pt]{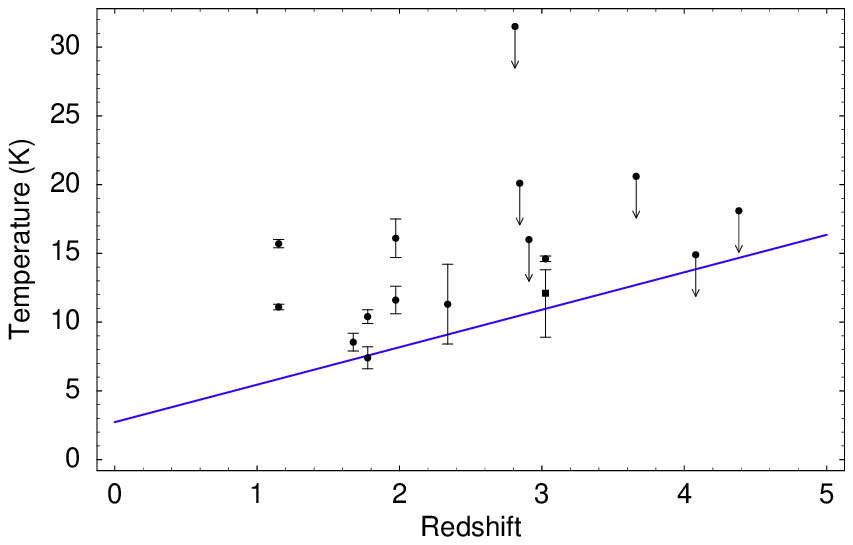}}
  \caption{Fine-structure excitation temperatures derived in this and in
    previous studies \citep{ASongailaCHR1994, LLuSBCV1996, JGeBB1997,
    KRothJBauer1999, RSrianandPL2000, PMolaroLDO2002}. The square dot
    marks the CMBR temperature derived by \cite{PMolaroLDO2002}, the solid
    line represents the prediction of the standard Friedmann cosmology}

  \label{fg:T}
\end{figure}

\begin{table}
  \caption{Excitation temperatures of the ground state fine-structure
    levels in \ion{C}{i}. The indices denote the $J$ quantum numbers
    of the levels considered}

  \begin{tabular}{llll}
    \hline
    Absorber\rule[-6pt]{0pt}{18pt}
      & $T_{01}$ (K) & $T_{02}$ (K) & $T_{12}$ (K) \\
    \hline
    Major\rule{0pt}{12pt}
      & $15.7\pm0.3$ & $18.5\pm0.3$ & $20.9\pm0.5$ \\
    Minor
      & $11.1\pm0.2$ & $15.0\pm0.6$ & $19.2\pm1.6$ \\
    \hline
  \end{tabular}

  \label{tb:tdata}
\end{table}

The fine-structure excitation temperatures derived in this and in previous
studies considering higher redshifts are compared in Fig.~\ref{fg:T}.

\section{Conclusions} 

Our theoretical calculations of the relative population of the ground
state fine-structure levels in \ion{C}{i} clearly demonstrate that the
CMBR is only a minor source of the observed fine-structure excitation. The
ratios of the observed column densities suggest that the \ion{C}{i}
absorbing medium is either very dense or exposed to very intense UV
radiation. The upper limit on the local UV energy density is 100 times the
galactic UV energy density, while the upper limit on the \ion{H}{i} number
density is 110~cm$^{-3}$. Whether flourescence induced by the local UV
radiation field or collisional excitation by hydrogen atoms is the more
important process cannot be concluded yet.

We also observe absorption lines of molecular hydrogen associated with the
DLA system. The observed rotational population is strongly inversed,
indicating not only collisonal but also radiative excitation. Therefore,
the kinetic temperature of the absorbing medium can only be determined in
a multilevel population analysis. This analysis is in progress and will
possibly unravel the physical processes giving rise to the observed
fine-structure population.

\begin{acknowledgements}
  We kindly acknowledge A.~I.~Silva and S.~M.~Viegas for providing their
  PopRatio program package for the public. We also thank N.~Hansen for
  sending us a manuscript and for providing us with helpful detailed
  information about his evolution strategy. This research has been
  supported by the Verbundforschung of the BMBF/DLR under Grant
  No.~50~OR~9911~1.
\end{acknowledgements}

\bibliographystyle{aa}
\bibliography{aa,astron}

\end{document}